\documentclass[
preprint,
prd,
aps,
amsmath,
amssymb,
]{revtex4-1}

\def\Tr{\mbox{Tr}\,}

\usepackage{graphicx}
\usepackage{float}
\usepackage{subfloat}
\usepackage{subfigure}

\usepackage{hyperref}

\begin{document}

\title{Alignment, reverse alignment, and wrong sign Yukawa couplings 
	in two Higgs doublet models} 

\author{Ambalika Biswas}\email{ambalika12t@bose.res.in}
\author{Amitabha Lahiri}\email{amitabha@bose.res.in}
\affiliation{S.~N.~Bose National Centre For Basic Sciences\\
Block JD, Sector III, Salt Lake, Kolkata 700106, INDIA}
\date{\today}

\begin{abstract}
	We consider two Higgs doublet models with a softly broken 
U(1) symmetry, for various limiting values of the scalar 
mixing angles $\alpha$ and $\beta$\,. These correspond to
the Standard Model Higgs particle being the lighter CP-even 
scalar (alignment) or the heavier CP-even scalar (reverse alignment),
and also the limit in which some of the Yukawa couplings 
of this particle are of the opposite sign from the vector boson couplings 
(wrong sign). In these limits we impose a criterion for naturalness
by demanding that quadratic divergences cancel at one loop. We
plot the allowed masses of the remaining physical scalars 
based on naturalness, stability, perturbative unitarity and
constraints coming from the $\rho$ parameter.  We also calculate the 
$h\to \gamma\gamma$ decay rate in the wrong sign limit.
\end{abstract}

\maketitle

\section{Introduction}
The discovery of a new boson in July 2012 by the ATLAS~\cite{Atlas} and 
CMS Collaborations~\cite{CMS} at the Large Hadron Collider (LHC) is a 
landmark in the history of Particle Physics. This scalar is most likely \textit{the}
Higgs boson which is the last missing block in the Standard Model (SM). 
Although it answers most of the questions concerning fundamental particles, 
the SM has a few shortcomings, thus encouraging a search for theories
beyond the Standard Model. Among the inadequacies are the lack of  clear 
answers on the questions of the origins of neutrino mass
and dark matter. It also cannot provide the observed
matter-antimatter asymmetry of the universe.
 
One of the simplest ways to go beyond the SM is by extending the scalar 
sector. This of course affects the $\rho$ parameter, whose deviation from 
the tree level value of unity is a measure of new physics. The general 
expression for the tree level $\rho$ parameter 
for an SU(2)$\times$U(1) gauge theory with $N$ scalar multiplets 
is~\cite{Langacker:1980js}
\begin{equation}
\rho\equiv
\frac{m^2_W}{\cos^2\theta_W\, m^2_Z} =
\frac{\sum_{{i}=1}^{N}\left[T_{{i}}(T_{{i}}+1)-\frac{1}{4} 
	{Y_{i}^{2}}\right]v_i^2}
{\frac{1}{2}\sum_{{i}=1}^{N}Y_{{i}}^{2}v_i^2}\,,
\end{equation}
where $T_{{i}}$ and $Y_{{i}}$ denote the weak isospin 
and hypercharge of the $i^{th}$ scalar multiplet respectively, and 
$v_{{i}}$ is the vacuum expectation value (vev) of the neutral component of 
that multiplet. If the scalar sector contains only SU(2) singlets 
with $Y = 0$ and doublets with $Y = \pm1$\,, then 
$\rho = 1$ is automatically satisfied without 
requiring any fine tuning among the vevs. This conforms with 
the experimental value of $\rho$, which is very close to
unity~\cite{Agashe:2014kda}. We therefore confine our discussions to the 
doublet extensions,  specifically the two Higgs-doublet models
(2HDMs)~\cite{Branco:2011iw}, which have received a lot of attention 
mainly because the Type II 2HDM arises as part of minimal supersymmetry.

In this paper we consider the restrictions imposed on the scalar masses
by a criterion of naturalness, embodied in the Veltman
conditions, in various limits of 2HDMs of all types. The alignment limit 
and the reverse alignment limit are two scenarios in which the lighter 
and the heavier CP-even neutral scalar, respectively, correspond to the observed 
Higgs particle. We also consider the cases where these occur in
conjunction with the wrong sign limit, in which the Yukawa coupling
of at least one type of fermion is of the opposite sign as the vector
coupling. Using the naturalness conditions we analyze the parameter space
of masses of scalars in 2HDMs of different types. The parameter space 
is further restricted by constraints arising from the $\rho$-parameter, 
global stability of the scalar potential, and requirement of perturbative unitarity. 
Section II gives a brief review of 2HDM. Sections III and IV deal with 
various limits of two Higgs doublet models and their permutations. 
In section V we calculate the Higgs-diphoton decay width for one of 
the scenarios and section VI concludes with a discussion of the results.

\section{Brief review of 2HDMs}
We will work with the  scalar potential~\cite{Lee:1973iz,
  Gunion:1989we} considered under the imposition of a U(1) symmetry which forbids 
flavor-changing neutral currents (FCNCs),
\begin{eqnarray}
V &=&
\lambda_{1}\left(|\Phi_{1}|^2 
- \frac{v_{1}^{2}}{2}\right)^{2} + 
\lambda_{2}\left(|\Phi_{2}|^2 
- \frac{v_{2}^{2}}{2}\right)^{2}     
 \nonumber \\
 & &\quad 
+\lambda_3\left(|\Phi_1|^2 + |\Phi_2|^2
-\frac{v_{1}^{2} +  v_{2}^{2}}{2}\right)^{2}  
\nonumber \\ 
&& \quad
+\lambda_{4}\left(|\Phi_{1}|^2 |\Phi_{2}|^2 
- |\Phi_{1}^{\dagger}\Phi_{2}|^2\right) 
 \nonumber \\
 & &\quad 
 + \lambda_5\left|\Phi_1^\dagger\Phi_2 -  \frac{v_1v_2}{2}\right|^2\,,
\label{2HDM.potential}
\end{eqnarray}
with real $\lambda_i$.  This potential is
invariant under the symmetry 
 $\Phi_1 \to e^{i\theta}\Phi_1\,, \Phi_2 \to \Phi_2\,,$
except for a soft breaking term $\lambda_5 v_1v_2
\Re(\Phi_{1}^{\dagger} \Phi_{2})\,.$ Additional dimension-4 terms,
including one allowed by a softly broken $Z_2$ symmetry~\cite{Gunion:1992hs}
are also set to zero by this U(1) symmetry. This is the 
same U(1) symmetry which prevents FCNC by having left-
and right-handed fermions transform differently under it,
leading to the four types of 2HDMs.


The  scalar doublets are parametrized as
\begin{equation}
\Phi_{i}=
\left(
\begin{array}{c}
w_{i}^{+}(x)\\
\frac{v_{i}+h_{i}(x)+iz_{i}(x)}{\sqrt{2}}
\end{array}
\right)\,, \qquad i=1,2
\label{doublet.vev}
\end{equation}
where the VEVs $v_i$ may be taken to be real and positive without
any loss of generality.  Three of these fields get ``eaten'' by the
$W^{\pm}$ and $Z^{0}$ gauge bosons; the remaining five are physical
scalar fields. There is a pair of charged scalars denoted
by $\xi^{\pm}$, two neutral CP-even scalars $H$ and $h$\,,
and one CP-odd pseudoscalar denoted by
$A$. The two CP-even scalars have distinct masses, and 
$m_h < m_H\,.$
 With
\begin{equation}
\tan\beta =\frac{v_{2}}{v_{1}}\,, 
\label{tanbeta.def}
\end{equation}
%
the scalar fields are given by the combinations
\begin{equation}
\left(
\begin{array}{c}
\omega^{\pm}\\
\xi^{\pm}
\end{array}
\right)=\left(
\begin{array}{rcl}
c_{\beta}& s_{\beta}\\
-s_{\beta}& c_{\beta}
\end{array}
\right) \left(
\begin{array}{c}
w_{1}^{\pm}\\
w_{2}^{\pm}
\end{array}
\right),
\label{redef.charged}
\end{equation}
\begin{equation}
\left(
\begin{array}{c}
\zeta\\
A
\end{array}
\right)=\left(
\begin{array}{rcl}
c_{\beta}& s_{\beta}\\
-s_{\beta}& c_{\beta}
\end{array}
\right) \left(
\begin{array}{c}
z_{1}\\
z_{2}
\end{array}
\right),
\label{redef.axial}
\end{equation}
\begin{equation}
\left(
\begin{array}{c}
H\\
h
\end{array}
\right)=\left(
\begin{array}{rcl}
c_{\alpha}& s_{\alpha}\\
-s_{\alpha}& c_{\alpha}
\end{array}
\right) \left(
\begin{array}{c}
h_{1}\\
h_{2}
\end{array}
\right),
\label{redef.higgs}
\end{equation}
where  $c_{\alpha}\equiv \cos\alpha\,,$ etc. We will assume, without loss 
of generality, that $0\leq\beta\leq\frac{\pi}{2}$\,, and 
$-\frac{\pi}{2}\leq\alpha\leq\frac{\pi}{2}$\,.

The quartic couplings are related to the physical Higgs masses 
by~\cite{Kanemura:1993hm, Akeroyd:2000wc}:
\begin{eqnarray}
	\lambda_{1} &=& \frac{1}{2v^{2} c^2_\beta}\left[c^2_{\alpha} m_{H}^{2}+s^2_{\alpha} m_{h}^{2}-\frac{s_{\alpha}c_{\alpha}}{\tan\beta}(m_{H}^{2}-m_{h}^{2})\right]
	-\frac{\lambda_{5}}{4}(\tan^{2}\beta-1)\,, \label{mass.lambda1}\\  
	\lambda_{2} &=& \frac{1}{2v^{2}s^{2}_{\beta}}\left[s^2_{\alpha} m_{H}^{2}+c^2_{\alpha} m_{h}^{2}-s_{\alpha}c_{\alpha}\tan\beta(m_{H}^{2}-m_{h}^{2})\right]
	-\frac{\lambda_{5}}{4}\left(\frac{1}{\tan^{2}\beta} -1\right)\,,
	\label{mass.lambda2}\\ 
	\lambda_{3} &=& \frac{1}{2v^{2}}\frac{s_{\alpha}c_{\alpha}}{s_{\beta}c_{\beta}}(m_{H}^{2}-m_{h}^{2})
	-\frac{\lambda_{5}}{4}\,,\label{mass.lambda3}\\  
	\lambda_{4} &=& \frac{2}{v^{2}}m_{\xi}^{2}\,,
	\label{mass.lambda4}\\
	\lambda_5 &=& \frac{2}{v^{2}}m_{A}^{2}\,.
	\label{mass.lambda5}
\end{eqnarray}

Let us now turn our attention to the fermion couplings. 
The scalar doublets couple to the fermions in the theory via the Yukawa
 Lagrangian
\begin{equation}
{\cal L}_{Y}= \sum_{i=1,2}
\left[- \bar{l}_L\Phi_iG_{e}^i e_R 
- \bar{Q}_L \tilde{\Phi}_{i}
G_{u}^{i}u_{R}
- \bar{Q}_{L}\Phi_{i}G_{d}^{i} d_{R} + h.c.\right]\,.
\end{equation}
Here $l_L\,, Q_L$ are 3-vectors of isodoublets in the space of
generations, $e_R\,, u_R\,, d_R$ are 3-vectors of singlets, $G^1_e$
etc. are complex $3\times 3$ matrices in generation space
containing the Yukawa coupling constants, and
$\tilde\Phi_i=i\tau_2\Phi_i^*\,.$

When the fermions are in mass eigenstates, the Yukawa
matrices are automatically diagonal if there is only one Higgs
doublet as in the Standard Model. But in the presence of a 
second scalar doublet, the two
Yukawa matrices will not be simultaneously diagonalizable in
general. Thus the Yukawa couplings will not be flavor
diagonal, and neutral Higgs scalars will mediate 
FCNCs~\cite{Branco:1996bq, Botella:2014ska, 
	Bhattacharyya:2014nja}.  
The necessary and sufficient condition 
for the absence of FCNCs at tree level is that all fermions of a given 
charge and helicity transform according to
the same irreducible representation of SU(2), corresponding to the
same eigenvalue of $T_{3}\,,$ and that a basis exists in which they
receive their contributions in the mass matrix from a single
source~\cite{Glashow:1976nt, Paschos:1976ay}.

For the fermions of the Standard Model,
this theorem implies that all right-handed singlets of a given
charge must couple to the same Higgs doublet. This can be 
ensured by using the global U(1) symmetry mentioned earlier, 
which generalizes a $Z_2$ symmetry more commonly employed 
for this purpose. The left handed fermion doublets remain 
unchanged under this
symmetry, $Q_L \to Q_L\,, l_L \to l_L\,.$ The transformations of
right handed fermion singlets determine the type of 2HDM. There are
four such possibilities, which may be identified by the
right-handed fields which transform under the U(1): type I (none),
type II ($d_{R}\rightarrow e^{-i\theta}d_{R}\,, e_{R}\rightarrow
e^{-i\theta}e_{R}$)\,, lepton specific ($e_{R}\rightarrow
e^{-i\theta}e_{R}$)\,, flipped ($d_{R}\rightarrow
e^{-i\theta}d_{R}$)\,. 

The scalar masses get quadratically divergent contributions which 
require very large fine-tuning of parameters. We will impose a 
criterion of naturalness on the scalar masses, viz., the cancellation
of these quadratic divergences. This gives rise to four mass relations, 
which we may call the Veltman conditions for the 2HDMs being 
considered~\cite{Newton:1993xc},
%
\begin{eqnarray}
2\Tr G_{e}^{1}G_{e}^{1\dagger} + 6\Tr G_{u}^{1\dagger}G_{u}^{1} 
+ 6\Tr G_{d}^{1}G_{d}^{1\dagger} &=& 
\frac{9}{4}g^{2}+\frac{3}{4}g^{\prime 2}+6\lambda_{1}
+10\lambda_{3}+\lambda_{4} + \lambda_5 \,, 
\label{VC.vc1}\\
2\Tr G_{e}^{2}G_{e}^{2\dagger} + 6\Tr G_{u}^{2\dagger}G_{u}^{2}
+ 6\Tr G_{d}^{2}G_{d}^{2\dagger} &=&
\frac{9}{4}g^{2}+\frac{3}{4}g^{\prime 2}+6\lambda_{2}
+10\lambda_{3}+\lambda_{4}+ \lambda_5\,,
\label{VC.vc2} \\
2\Tr G_{e}^{1}G_{e}^{2\dagger} + 6\Tr G_{u}^{1\dagger}G_{u}^{2}
+ 6\Tr G_{d}^{1}G_{d}^{2\dagger} &=&0\,,
\label{VC.vc3}
\end{eqnarray}
%
and another one which is the complex conjugate of the third equation.
Here $g, g'$ are the $SU(2)$ and $U(1)_Y$ coupling constants, respectively.

The fermion mass matrix is diagonalized by independent unitary
transformations on the left and right-handed fermion fields. In any
of the 2HDMs, the U(1) symmetry implies that either $G_{1f}$ 
or $G_{2f}$ must vanish for each fermion type $f\,.$ 
For example, in the Type II model $\Phi_{1}$ couples to
down-type quarks and charged leptons,  while $\Phi_{2}$ couples to
up-type quarks, so $G_{2e}= G_{2d}= G_{1u}=0\,.$ Thus
Eq.~(\ref{VC.vc3}) is automatically satisfied in each 2HDM, and 
 the relevant mass relations come from the first two equations above.
The non-vanishing Yukawa matrices are related to the fermion masses
by
\begin{eqnarray}
\Tr[G_{1f}^{\dagger}G_{1f}] &=&\frac{2}{v^2 \cos^{2}\beta} \sum
m_f^2\,, \label{Yukawa.1}\\  
\Tr[G_{2f}^{\dagger}G_{2f}] &=&\frac{2}{v^2\sin^{2}\beta} \sum
m_f^2\,, \label{Yukawa.2} 
\end{eqnarray}
where $f$ stands for charged leptons, up-type
quarks, or down-type quarks, and the sum is taken over generations.
These and the scalar mass relations of Eqs.~(\ref{mass.lambda1}) 
--~(\ref{mass.lambda5}) allow us to write the Veltman
conditions in terms of the physical masses of particles. 

There are some additional conditions on the parameters which further
constrain the scalar masses. One is the pertubativity condition,
which puts a constraint on the quartic coupling constants, 
$\lambda_{i}\leq 4\pi$~\cite{Kanemura:1999xf}. Another set
 comes from the condition that
the potential is bounded from below. This was examined for 
more general potentials in 2HDM under U(1) symmetry 
in~\cite{Sher:1988mj, Gunion:2002zf}, and for the potential given
in Eq.~(\ref{2HDM.potential}) these conditions become 
\begin{eqnarray}
\lambda_{1}&+&\lambda_{3}>0\,,\label{S1}\\
\lambda_{2}&+&\lambda_{3}>0\,,\label{S2}\\
2\lambda_{3}+\lambda_{4}&+&2\sqrt{(\lambda_{1}+\lambda_{3})(\lambda_{2}
	+\lambda_{3})}>0\,,\label{S3}\\
2\lambda_{3}+\lambda_{5}&+&2\sqrt{(\lambda_{1}+\lambda_{3})(\lambda_{2}
	+\lambda_{3})}>0\,.\label{S4}
\end{eqnarray}
These conditions put lower bounds on the above combinations of quartic couplings, but
there are also upper bounds on these couplings arising from the considerations of 
perturbative unitarity~\cite{Lee:1977eg}. These conditions are
\begin{eqnarray}
\vert 2\lambda_{3}-\lambda_{4}+2\lambda_{5}\vert\leq 16\pi\,,\label{PU1}\\
\vert 2\lambda_{3}+\lambda_{4}\vert\leq 16\pi\,,\label{PU2}\\
\vert 2\lambda_{3}+\lambda_{5}\vert\leq 16\pi\,,\label{PU3}\\
\vert 2\lambda_{3}+2\lambda_{4}-\lambda_{5}\vert\leq 16\pi\,,\label{PU4}\\
\vert 3(\lambda_{1}+\lambda_{2}+2\lambda_{3})\pm \sqrt{9(\lambda_{1}
	-\lambda_{2})^{2}+(4\lambda_{3}+\lambda_{4}
	+\lambda_{5})^{2}}\vert \leq 16\pi\,,\label{PU5}\\
\vert (\lambda_{1}+\lambda_{2}+2\lambda_{3})\pm \sqrt{(\lambda_{1}
	-\lambda_{2})^{2}+(\lambda_{4}-\lambda_{5})^{2}}\vert \leq 16\pi\,,\label{PU6}\\
\vert (\lambda_{1}+\lambda_{2}+2\lambda_{3})\pm (\lambda_{1}-\lambda_{2})\vert \leq 16\pi\,.
\label{PU7}
\end{eqnarray}

There is another condition that we need to take into account when
we calculate bounds on the scalar masses. 
%
%
The oblique electroweak correction $T$\,, which measures deviations from 
the standard model due to new physics, is related to the deviation
of the $\rho$ parameter from its SM value of unity by
\begin{equation}
\delta\rho \equiv \rho - 1 = \alpha T\,,
\end{equation}
where $\alpha = e^2/4\pi$ is the fine structure constant. 
The effect of the general 2HDM on the $\rho$ parameter is known 
to be~\cite{Grimus:2007if, Kanemura:2011sj}
\begin{align}
\delta\rho = &\,\frac{g^{2}}{64\pi^{2}m_{w}^{2}} \Big( F(m_{\xi}^{2},m_{A}^{2})
+\sin^{2}(\beta-\alpha)F(m_{\xi}^{2},m_{H}^{2})
+\cos^{2}(\beta-\alpha)F(m_{\xi}^{2},m_{h}^{2})\nonumber\\
&-\sin^{2}(\beta-\alpha)F(m_{A}^{2},m_{H}^{2})-\cos^{2}(\beta-\alpha)F(m_{A}^{2},m_{h}^{2})\nonumber\\
&+3\cos^{2}(\beta-\alpha)\left[F(m_{Z}^{2},m_{H}^{2})-F(m_{W}^{2},m_{H}^{2})\right]\nonumber\\
&+3\sin^{2}(\beta-\alpha)\left[F(m_{Z}^{2},m_{h}^{2})-F(m_{W}^{2},m_{h}^{2})\right]\nonumber\\
&-3\left[F(m_{Z}^{2},m_{h_{SM}}^{2})-F(m_{W}^{2},m_{h_{SM}}^{2})\right]\Big)\label{rho}\,,
\end{align}
where $F(x, y)$ is a function of two non-negative arguments ${x}$ 
and ${y}$\,, symmetrical under the exchange of the arguments 
and vanishes only if ${x=y}$. The function has the property that 
it grows linearly with  $\max({x,y}$), i.e., quadratically 
with the heaviest scalar mass when that mass becomes very large. 
The current experimental bound on the total new physics contribution 
to $\rho$ is given by $\delta\rho=-0.00011$~\cite{Agashe:2014kda}.

\section{Limits of 2HDMs}
In order to relate a 2HDM to the Higgs sector of the Standard Model,
we need to identify some combination of the neutral scalar particles 
in the theory as the observed Higgs particle. This can be done in 
several ways, by considering different combinations of the angles 
$\alpha$ and $\beta$\,. Is this section we will consider the different limits
for which part of the 2HDM matches the Standard Model, and calculate the 
allowed range of masses for the additional scalars. 

A crucial parameter of the 2HDMs is $\tan\beta\,.$ Its value is
larger than one, based on constraints coming from $Z\to
b\bar b$ and $B_q \bar B_q$ mixing~\cite{Arhrib:2009hc}. A large 
$\tan\beta$ is suggested by muon $g-2$ in lepton specific 
2HDM~\cite{Cao:2009as}, by using $b\to s\gamma$ in type I 
and flipped models~\cite{Park:2006gk}, which also suppresses 
the $t\rightarrow bH^{+}$ branching ratio
to a rough agreement with 95$\%$ CL limits from the
light charged Higgs searches at the LHC~\cite{Aad:2012tj,
	Chatrchyan:2012vca}. We will assume that $\tan\beta$ is 
large, and certainly larger than unity, specific values will be 
considered for the plots as needed.

\subsection{Alignment Limit}
If we rotated the neutral $(h_{1}\,, h_{2})$ doublet by the angle $\beta$, 
\begin{equation}
\left(
\begin{array}{c}
H^{0}\\
R
\end{array}
\right) = \left(
\begin{array}{rcl}
c_{\beta}& s_{\beta}\\
-s_{\beta}& c_{\beta}
\end{array}
\right) \left(
\begin{array}{c}
h_{1}\\
h_{2}
\end{array}
\right),
\label{Hbeta}
\end{equation}
we would find that $H^{0}$ has exactly the Standard Model Higgs
couplings with the fermions and gauge bosons~\cite{Branco:1996bq, 
	Gunion:2002zf}.
The physical scalar $h$ is related to $H^{0}$ and $R$ via 
\begin{equation}
h=\sin(\beta-\alpha)H^{0}+ \cos(\beta-\alpha)R\,.
\label{decoupling.1}
\end{equation}
Thus in order for $h$ to be the Higgs boson of the Standard Model,
we require $\sin(\beta-\alpha)\approx 1\,,$ which has been called the 
SM-like or alignment limit~\cite{Ferreira:2014naa}. 

There remain three unknown mass parameters, namely
$m_{H}, m_{\xi}$ and $m_{A}$, which span the parameter space. By fixing
$\tan\beta$ at some specific value, we can use the Veltman conditions
to plot the accessible region of the $m_H - m_\xi$ plane corresponding 
to the allowed range of values for $m_A\,.$ On the other hand, constraints
from perturbative unitarity and the oblique correction $T$ also restrict the 
accessible region on this plane. The intersection of all these regions
provides the allowed ranges for $m_H$ and $m_\xi$\,.

The mass ranges were studied for the alignment limit in~\cite{Biswas:2014uba},
where it was found that if we set $m_h = 125$ GeV, and allowed
$m_A$ to run over its entire range of $0<m_A \lesssim 617$ GeV as determined
by the condition of perturbativity, the two unknown masses $m_H$ and $m_\xi$
became restricted to ranges of 550 GeV $ \lesssim  m_\xi \lesssim$ 700 GeV,
450 GeV $\lesssim m_H \lesssim$ 620 GeV. The value of $\tan\beta$ used in
these calculations was $\tan\beta = 5\,,$ and it was also found that a higher value 
of $\tan\beta$ pushed the ranges to higher values and also made them narrower.
These mass ranges are in agreement with 
bounds found by analysing experimental data~\cite{Kanemura:2014dea}.

\subsection{Reverse Alignment Limit}
Let us rearrange the equations described in the previous section. 
Using Eqs.~(\ref{redef.higgs}) and~(\ref{Hbeta}) we obtain $H$ in terms of 
$H^{0}$ and $R$\,,
\begin{equation}
H=H^{0}\cos(\beta-\alpha)-R\sin(\beta-\alpha)
\end{equation}
Had $H$ been the SM-like Higgs boson, it would have to resemble the properties of $H^{0}\,,$ 
and for that $\beta$ would have to approximately equal $\alpha$ or $\pi +\alpha$.
The ultimate results with $\beta \approx \alpha$ and $\beta \approx \pi + \alpha$ are identical,
so in what follows we will work with $\beta \approx \alpha$ and call it the \textit{Reverse Alignment 
	Limit}.
 
Eqs.~(\ref{mass.lambda1}-\ref{mass.lambda5}) become, in the reverse alignment limit,
%
 \begin{eqnarray}
 \lambda_{1} &=& \frac{m_{h}^{2}}{2v^{2}}(\tan^{2}\beta+1)-\frac{\lambda_{5}}{4}(\tan^{2}\beta-1)\,,
 \label{RAL.lambda1}\\
 \lambda_{2} &=& \frac{m_{h}^{2}}{2v^{2}}(\cot^{2}\beta+1)-\frac{\lambda_{5}}{4}(\cot^{2}\beta-1)\,,
 \label{RAL.lambda2}\\
 \lambda_{3} &=& \frac{1}{2v^{2}}(m_{H}^{2}-m_{h}^{2})-\frac{\lambda_{5}}{4}\,,
 \label{RAL.lambda3}\\
\lambda_{4} &=& \frac{2}{v^{2}}m_{\xi}^{2}\,,
 \label{RAL.lambda4}\\
\lambda_{5} &=& \frac{2}{v^{2}}m_{A}^{2}\,. \label{RAL.lambda5} 
 \end{eqnarray}
Let us write the Veltman conditions defined in Eqs.~(\ref{VC.vc1})
and~(\ref{VC.vc2}) using the above equations.  We will write the
equations explicitly for one case, that of the Type II 2HDM, for which 
the two Veltman conditions read, in the reverse alignment limit, 
\begin{eqnarray}
m_{h}^{2}\left(3\tan^{2}\beta - 2\right) 
+ 2 m_{\xi}^{2}
= && 4\left[ \sum m_e^2 + 3 \sum m_d^2\right]\sec^{2}\beta
- 6M_W^{2} - 3M_Z^{2}
- 5 m_{H}^{2} + \lambda_{5}\frac{3v^{2}}{2}\tan^{2}\beta
\,,
\label{typeII.VC1.RAL} \\
m_{h}^{2}\left(3\cot^{2}\beta - 2\right) 
+ 2 m_{\xi}^{2}
= && 12 \sum m_u^2 \csc^{2}\beta - 6M_W^{2} - 3M_Z^{2}
- 5 m_{H}^{2} + \lambda_{5}\frac{3v^{2}}{2}\cot^{2}\beta
\,.   
\label{typeII.VC2.RAL}
\end{eqnarray}
\begin{figure}[htbp]
	\subfigure[]{
		\includegraphics[height=0.3\columnwidth, width = 0.45\columnwidth]{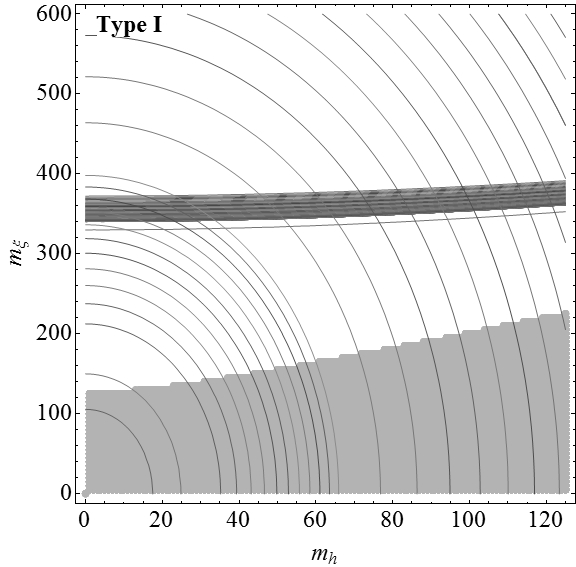} 
		\label{typeI.RAL}}
	\subfigure[]{
		\includegraphics[height=0.3\columnwidth, width = 0.45\columnwidth]{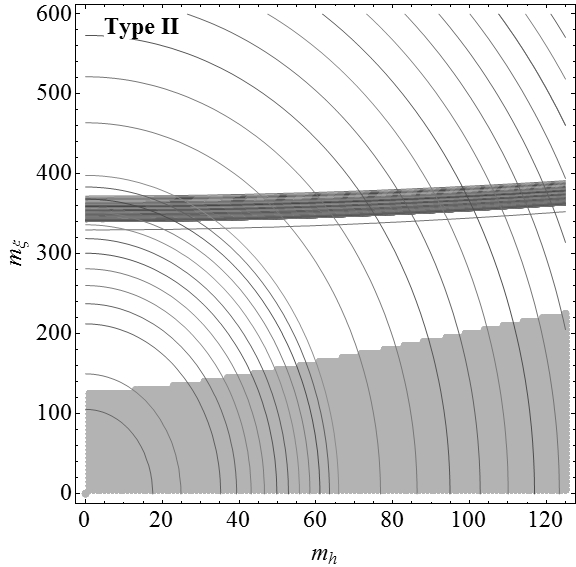}
		\label{type2.RAL}}
	\subfigure[]{
		\includegraphics[height=0.3\columnwidth, width = 0.45\columnwidth]{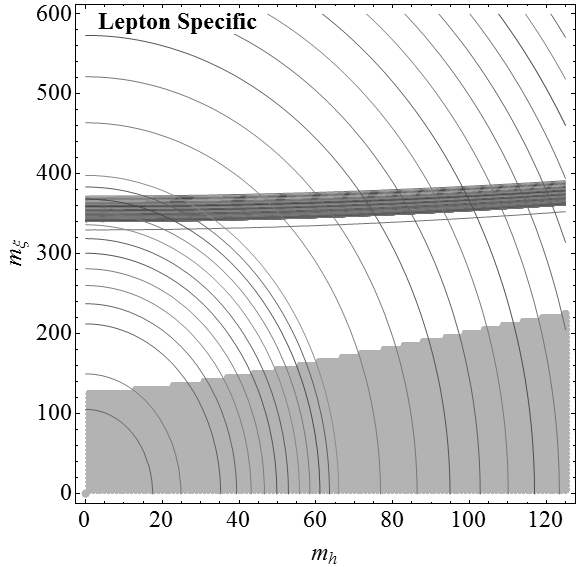}
		\label{LS.RAL}}
	\subfigure[]{
		\includegraphics[height=0.3\columnwidth, width = 0.45\columnwidth]{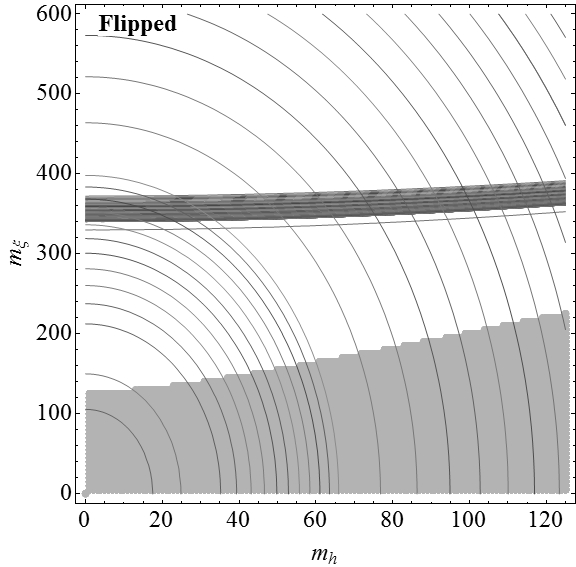}
		\label{Flipped.RAL}}
	\caption{Allowed mass range (in GeV) for the charged Higgs and 
		the light CP even Higgs in \textit{Reverse alignment limit} for (a) type I (b) type II (c) lepton specific and
		(d) flipped 2HDM 
		for $\vert\lambda_{5}
		\vert \leq 4\pi$ and $\tan \beta = 5\,.$}  
	\label{fig.result.RDC}
\end{figure}
We have plotted the above equalities on the $m_h - m_\xi$ plane 
for several values of $\lambda_5$ for a fixed value of $\tan\beta$
and with $m_H = 125$ GeV, with $m_h\leq m_H$\,. On the same plane, 
we have also plotted the region allowed by stability, 
perturbative unitarity, and constraints from  $\delta\rho\,.$ 
The conditions of stability and perturbative unitarity, Eq.~(\ref{S1}) 
-- Eq.~(\ref{PU7}), produce the following two inequalities in the 
reverse alignment limit relevant to this plot:
\begin{align}
		0  \leq \left(m_h^2 - m_A^2 \right) \left(\tan^2\beta + \cot^2\beta \right) 
		+ 2m_H^2 &\leq \frac{32\pi v^2}{3}\,,\label{gray.ineq1} \\
		\left|2 m_\xi^2 - m_h^2 - m_A^2 + m_H^2\right| &\leq 16\pi v^2\,.
\label{gray.ineq2}
\end{align}
These are analogous to similar inequalities found in~\cite{Biswas:2014uba}
in the alignment limit.

For $\tan\beta = 5\,,$ the plots for all four types of 2HDM are 
shown in Fig.~\ref{fig.result.RDC}. The gray region covers the points 
which satisfy the inequalities~(\ref{gray.ineq1}) and (\ref{gray.ineq2})
in addition to the constraints from $\delta\rho$, the first Veltman condition 
provides the curves (ellipses) which cross this region, and the 
second Veltman condition provides the nearly flat hyperbolas above
the gray region. 

As we can see from the plots in Fig.~\ref{fig.result.RDC}, there is no region on 
the $m_h-m_\xi$ plane where all the constraints are obeyed. In other words, if we 
insist on naturalness, as embodied by the Veltman conditions, the reverse alignment 
limit is not a valid limit for any of the 2HDMs, i.e. the observed Higgs particle 
cannot be the heavier CP-even neutral scalar in any of the 2HDMs. 

It should be mentioned here that allowed mass ranges of scalars in both the 
alignment limit and the reverse alignment limit were studied in~\cite{Coleppa:2013dya}. 
However, that paper considered an unbroken $Z_2$ symmetry, not a softly broken 
symmetry as we have considered. As a result the mass ranges of scalars, as well
as the allowed range of $\tan\beta$\, found in that paper, are different
from the ones we have found. 

\section{Wrong Sign Yukawa couplings}
The wrong-sign Yukawa coupling regime~\cite{Ferreira:2014naa, Ferreira:2014dya,
	 Ferreira:2014qda} is defined 
as the region of 2HDM parameter space in which at least one of the couplings of the 
SM-like Higgs to up-type and down-type quarks is opposite in sign to the corresponding 
coupling of SM-like Higgs to vectors bosons. 
This is to be contrasted with the Standard Model, where the couplings of $h_{SM}$ to $\bar{f}f$ 
and vector bosons are of the same sign. The \textit{wrong sign limit} 
needs to be considered in conjunction with either the alignment limit or the reverse 
alignment limit. We will now calculate the regions of parameter space when 
each of these two limits are combined with the wrong sign limit. 

The CP-even neutral scalars couple to the up-type and down-type quarks
in the various 2HDMs as shown in Table~\ref{Yukawa.Table}, with the 
SM couplings of the quarks to the SM Higgs field normalized to unity.
 \begin{table}[tbhps]
\begin{tabular}{||c|c|c|c|c||}
\hline
2HDMs & $h\bar{U}U\; $
&$\;  h\bar{D}D$ &$\; H\bar{U}U$ &$\; H\bar{D}D$ \\
\hline
Type I &$\;\frac{\cos\alpha}{\sin\beta} $ 
&$\; \frac{\cos\alpha}{\sin\beta} $  &$\;\frac{\sin\alpha}{\sin\beta} $
&$\; \frac{\sin\alpha}{\sin\beta} $\\
\hline 
Type II &$\;\frac{\cos\alpha}{\sin\beta} $ 
&$\; -\frac{\sin\alpha}{\cos\beta} $  &$\;\frac{\sin\alpha}{\sin\beta} $
&$\; \frac{\cos\alpha}{\cos\beta} $\\
\hline 
Lepton Specific &$\;\frac{\cos\alpha}{\sin\beta} $ 
&$\; \frac{\cos\alpha}{\sin\beta} $  &$\;\frac{\sin\alpha}{\sin\beta} $
&$\; \frac{\sin\alpha}{\sin\beta} $\\
\hline 
Flipped &$\;\frac{\cos\alpha}{\sin\beta} $ 
&$\; -\frac{\sin\alpha}{\cos\beta} $  &$\;\frac{\sin\alpha}{\sin\beta} $
&$\; \frac{\cos\alpha}{\cos\beta} $\\
\hline
\end{tabular}
\caption{Yukawa couplings for the different 2HDMs}
\label{Yukawa.Table}
\end{table}

\subsection{Wrong Sign and Reverse alignment limit}
Let us first consider the case of wrong sign Yukawa couplings in the 
reverse alignment limit. The heavier CP-even neutral scalar $H$ corresponds
to the SM Higgs in the reverse alignment limit, with a coupling 
to vector bosons which is $\cos(\beta - \alpha)$ times the corresponding 
SM value. In the convention where $\cos(\beta-\alpha)\geq 0 $, the 
$HVV$ couplings in the 2HDM are always non-negative. To analyze the
wrong-sign coupling regime, we write the 
 Yukawa couplings in the type-II and Flipped 2HDMs
in the following form:
\begin{eqnarray}
H\bar{D}D: \qquad \frac{\cos\alpha}{\cos\beta}&=&\cos(\beta+\alpha)
+\sin(\beta+\alpha)\tan\beta\,,
\label{HDD coupling}\\
H\bar{U}U: \qquad \frac{\sin\alpha}{\sin\beta}&=&-\cos(\beta+\alpha)
+\sin(\beta+\alpha)\cot\beta\,.  
\label{HUU coupling}
\end{eqnarray}

In the case when $\cos(\beta + \alpha) = -1$, the $H\bar{D}D$ coupling normalized to 
its SM value is equal to $-1$\,, whereas the normalized $H\bar{U}U$ coupling is +1\,. 
Thus in this case, when the reverse alignment limit is taken in conjunction with 
the wrong sign limit, we have $\alpha \approx \beta \approx \frac{\pi}{2}\,.$ It 
turns out there is 
no point on the $m_h - m_\xi$ plane which satisfies the Veltman conditions as
well as the bounds coming from unitarity, stability and the $\rho$-parameter. 
\begin{figure}[htbp]
\includegraphics[height=0.3\columnwidth, width = 0.45\columnwidth]{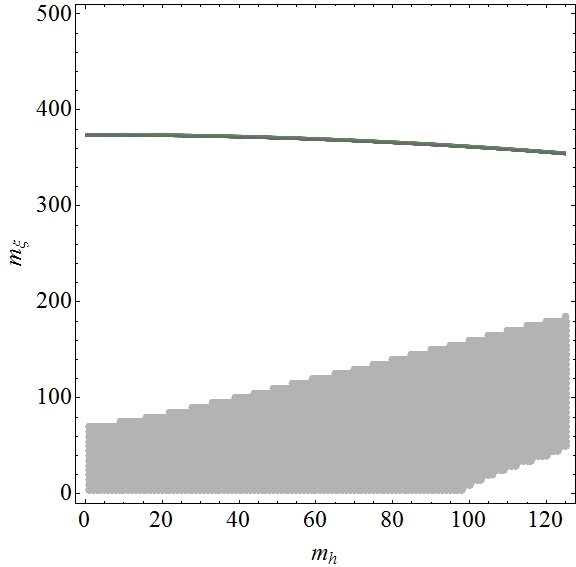}
\caption{Veltman conditions are not satisfied for any $(m_h, m_\xi)$ satisfying 
	unitarity and other bounds, in the reverse alignment limit with wrong sign 
	Yukawa couplings.}
\label{reversewrong.fig}
\end{figure}
In Fig.~\ref{reversewrong.fig} only the first Veltman condition has been plotted, and 
it does not cross the grey region corresponding to the bounds. The other Veltman condition 
does not show up in this picture at all, it is not satisfied for any point in this plot. 

On the other hand, in the case when $\cos(\beta + \alpha) = 1$, the $H\bar{U}U$ 
coupling normalized to its SM value is equal to $-1$, while the normalized 
$H\bar DD$ coupling is +1. In this limiting case, $\cos(\beta - \alpha) = 
\cos 2\beta$, which implies that the wrong-sign $H\bar{U}U$ couplings 
can only be achieved for $\tan\beta < 1$ for the type II and Flipped
2HDMs.

In the type-I and lepton specific 2HDMs, both the $H\bar{D}D$ and 
$H\bar{U}U$ couplings are given by Eq.~(\ref{HUU coupling}). 
Thus, for $\cos(\beta + \alpha) = 1$, both the normalized $H\bar{D}D$ 
and $H\bar{U}U$ couplings are equal to $-1$, 
which is only possible if $\tan\beta < 1$.

Since $\tan\beta > 1\,,$ we see that the wrong-sign Yukawa coupling
is incompatible with the reverse alignment limit in all of the four types
of 2HDMs. 




\subsection{Wrong sign in the Alignment limit}
Let us now look at what happens if some Yukawa couplings are of the 
wrong sign, in the alignment limit. In this case $h$ is the SM Higgs,
and its coupling to the vector bosons is $\sin(\beta - \alpha)$ times 
the corresponding SM value. Then in the convention where 
$\sin(\beta-\alpha)\geq 0 $, the $hVV$ couplings in the 2HDM 
are always non-negative. As in the previous case,
we write the type-II and Flipped Higgs-fermion Yukawa couplings,
normalized with respect to the Standard Model couplings, in the following form:
\begin{eqnarray}
h\bar{D}D: \qquad -\frac{\sin\alpha}{\cos\beta}&=& -\sin(\beta+\alpha)
+\cos(\beta+\alpha)\tan\beta\,,
\label{hDD coupling}\\
h\bar{U}U: \qquad \phantom{-}\frac{\cos\alpha}{\sin\beta}&=& \sin(\beta+\alpha)
+\cos(\beta+\alpha)\cot\beta\,.  
\label{hUU coupling}
\end{eqnarray}
In the case when $\sin(\beta + \alpha) = 1$, the $h\bar{D}D$ coupling normalized to 
its SM value is equal to $-1$\,, while the normalized $h\bar{U}U$ coupling is +1\,. 
Note that in this limiting case, $\sin(\beta - \alpha) = -\cos 2\beta$, which implies that the wrong-sign $h\bar{D}D$ Yukawa coupling can only be achieved for values of 
$\tan\beta > 1$. 

Likewise, in the case of $\sin(\beta + \alpha) = -1$, the $h\bar{U}U$ 
coupling normalized to its SM value is equal to $-1$\,, whereas the 
normalized $h\bar DD$ coupling is +1\,. Then $\sin(\beta - \alpha) 
= \cos 2\beta$, which implies 
that the wrong-sign $h\bar{U}U$ couplings can occur only if $\tan\beta < 1$. 
In the type-I and lepton specific 2HDM, both the $h\bar{D}D$ and $h\bar{U}U$ couplings are given by
Eq.~(\ref{hUU coupling}). Thus for $\sin(\beta + \alpha) = -1$, 
both the normalized $h\bar{D}D$ and $h\bar{U}U$ couplings are equal to $-1$, 
which is only possible if $\tan\beta < 1$.
Thus realistically only the $h\bar{D}D$ coupling of the type-II and flipped 2HDM 
can be of the wrong sign, since $\tan\beta > 1$.

Let us therefore consider a type II model with a wrong sign $h\bar{D}D$ coupling.
The wrong sign limit approaches the alignment limit for $\tan\beta\approx17$ 
as was displayed in~\cite{Ferreira:2014dya, Ferreira:2014qda} for the allowed parameter 
space of the type II CP-conserving 2HDM, based on the 8 TeV run  of the LHC.
For this model, we will plot the values of the pair $(m_H, m_\xi)\,$ allowed by
the naturalness conditions as well as the constraints imposed by perturbativity,
stability, tree-level unitarity, and the $\rho$ parameter. We will do this for
four different values of $\tan\beta\,$ around the `critical' value of 17.
By choosing a small enough $\alpha$ we can ensure that for all these
choices, both $\sin(\beta - \alpha)\approx 1$ and $\sin(\beta 
+ \alpha)\approx 1\,,$ 
as needed for the alignment limit and the wrong sign coupling.

\begin{figure}[htbp]
\subfigure[]{
 \includegraphics[height=0.3\columnwidth, width = 0.45\columnwidth]{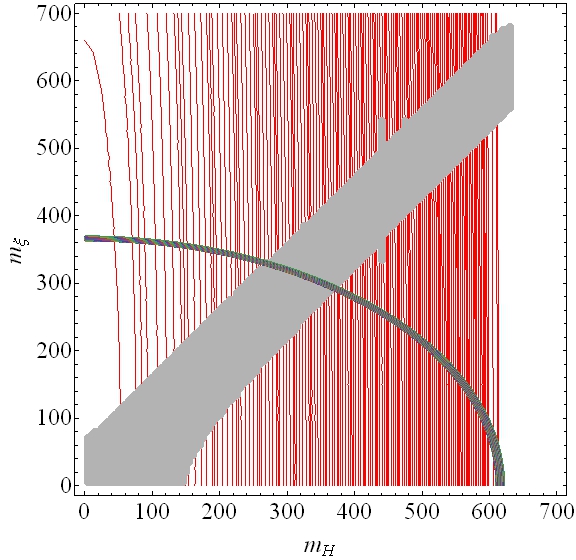} 
 \label{typeII.WSLAL.tan10}}
\subfigure[]{
 \includegraphics[height=0.3\columnwidth, width = 0.45\columnwidth]{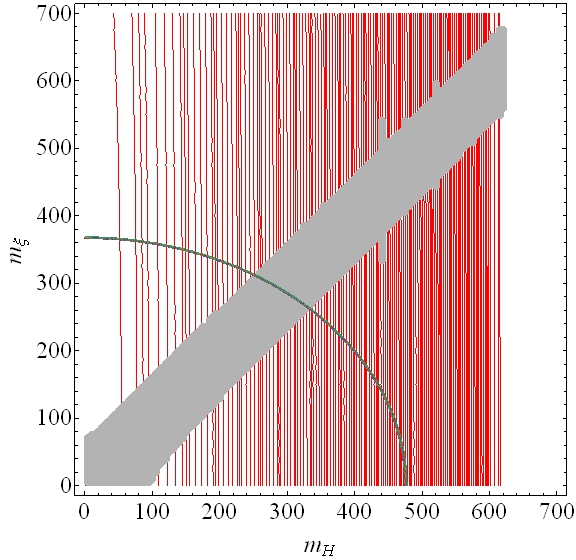} 
  \label{typeII.WSLAL.tan17}}
\subfigure[]{
 \includegraphics[height=0.3\columnwidth, width = 0.45\columnwidth]{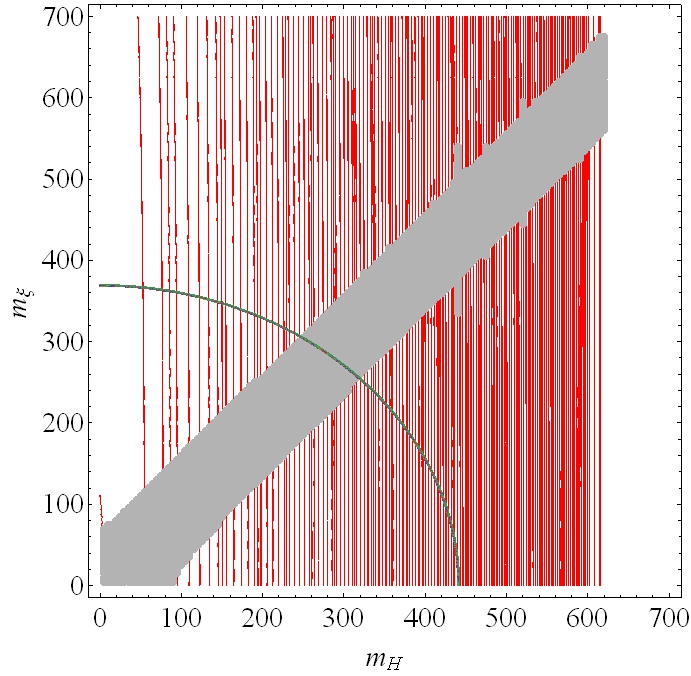}
 \label{typeII.WSLAL.tan20}}
\subfigure[]{
 \includegraphics[height=0.3\columnwidth, width = 0.45\columnwidth]{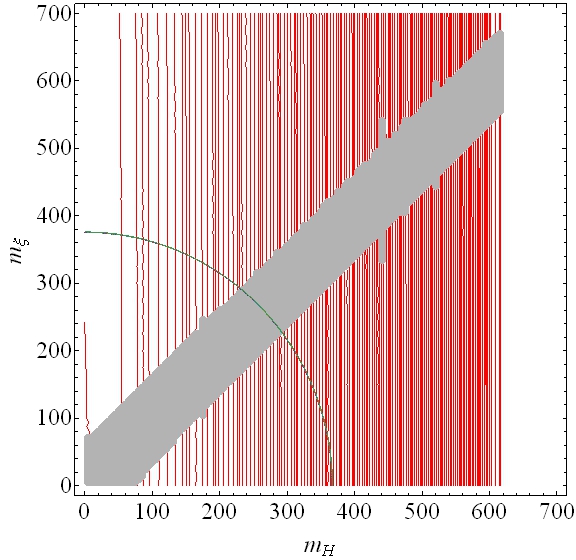}
 \label{typeII.WSLAL.tan30}}
\caption{Allowed mass range in GeV for the charged Higgs and 
the heavy CP even Higgs when approaching 
\textit{wrong sign and alignment} limits simultaneously for 
(a) $\tan \beta = 10$ (b)  $\tan \beta = 17$ (c)  $\tan \beta = 20$ and
(d) $\tan \beta = 30$
for $\vert\lambda_{5}
 \vert \leq 4\pi$ and Type II 2HDM.}  
\label{wslal.fig}
\end{figure}

In Fig.~\ref{wslal.fig} we have plotted the Veltman conditions on 
the $m_H-m_\xi$ plane for 
Type II 2HDM for the four choices of $\tan\beta$\,, for different values
of $m_A$ constrained by $|\lambda_{5}|\leq 4\pi$\,. This plots are further 
constrained by conditions coming from stability of the potential, 
perturbative unitarity, and experimental bounds on $\delta\rho$\,.
We have also taken $m_{h}=125$ GeV. One can estimate from the plots that 
for $\tan\beta=17\,$ that the range of $m_H$ is approximately 
(250, 330) GeV, and that of $m_\xi$ is approximately (260, 310) GeV. 
At higher values of $\tan\beta\,,$  both ranges become narrower and move 
down on the mass scale. 
%

\section{Modification of Higgs-diphoton decay width}
The $h\to \gamma\gamma$ decay channel is perhaps the most popular
channel for Higgs and related searches. 
The decay width can be enhanced or reduced in the 2HDMs due to loop effects.
In the alignment limit, the couplings of the lighter CP 
even neutral scalar ${h}$ to gauge bosons are identical to that
for the SM Higgs. Then the tree level decay widths of ${h}$ will 
be the same as for the SM Higgs. For loop induced decays, such as 
$ h\rightarrow \gamma\gamma $ and $ h\rightarrow Z\gamma\,,$ the 
contribution of the ${W}$ boson loop 
and the top loop diagrams are the same as in the SM. But there 
will  have some additional contributions 
due to the virtual charged scalars $ \xi^{\pm} $ in the loop. 
Thus the decay widths will be different from the SM in general.
Contributions from the fermion loops are the same in this case
as for the SM. 

On the other hand, suppose $h$ has wrong sign Yukawa couplings
to the down-type quarks. Then the bottom quarks will contribute 
with a relative negative sign in the loops, and the $h\to\gamma\gamma$ 
decay width will be different from the SM, as well as from 2HDMs in 
the usual alignment limit. 

The Higgs-diphoton decay width is calculated using the formula~\cite{Djouadi:2005gj}
\begin{equation}
\Gamma(h\rightarrow \gamma\gamma)=\frac{G_{\mu}\alpha^{2}m_{h}^{3}}{128\sqrt{2}\pi^{3}}
\left\vert\sum_{\textit{f}}N_{c}Q_{f}^{2}g_{hff}A_{1/2}^{h}(\tau_{f})+g_{hVV}A_{1}^{h}
(\tau_{W})+\frac{m_{W}^{2}\lambda_{h\xi^{+}\xi^{-}}}{2c_{W}^{2}M_{\xi^{\pm}}^{2}}
A_{0}^{h}(\tau_{\xi^{\pm}})\right\vert^{2}\,.
\label{decay_rate}
\end{equation}
In this equation, $N_c$ is the number color multiplicity, $Q_f$ is the charge of 
the fermion $f\,,$ $G_\mu$ is the Fermi constant, and the reduced couplings $g_{hff}$ 
and $g_{hVV}$ of the Higgs boson to 
fermions and $W$ bosons are $g_{\textit{htt}}=\dfrac{\cos\alpha}{\sin\beta}\,,  \,
g_{\textit{hbb}}=-\dfrac{\sin\alpha}{\cos\beta}\,$ and  $g_{\textit{hWW}}=\sin(\beta-\alpha)\,,$ 
while the trilinear $\lambda_{h\xi^{+}\xi^{-}}$ couplings to charged Higgs bosons is given by
\begin{eqnarray}
\lambda_{h\xi^{+}\xi^{-}} &=& \cos2\beta\sin(\beta+\alpha) + 2c^{2}_{W}\sin(\beta-\alpha)\\
&=& \lambda_{hAA}+ 2c^{2}_{W} g_{hVV}\,,
\end{eqnarray}
where $c_{W}=\cos\theta_{W}$\,, with $\theta_{W}$ being the Weinberg angle.
The decay rate does not depend on the type of the 2HDM.
  
The amplitudes $A_{i}$ at lowest order for the spin-1, spin-$\frac{1}{2}$ and 
spin-0 particle contributions are given by~\cite{Gunion:1989we}
\begin{eqnarray}
A^{\textit{h}}_{1/2} &=& -2\tau[1 +(1-\tau)f(\tau)]\\
A^{\textit{h}}_{1} &=& 2 +3 \tau +3\tau(2-\tau)f(\tau)\\
A^{\textit{h}}_{0} &=& \tau [1 -\tau f(\tau)]
\end{eqnarray}
in the case of the CP even Higgs boson $h$.

Here 
\begin{equation}
\tau_{x}=4m_{x}^{2}/m_{h}^{2}
\end{equation}
 and
 \begin{equation}
 f(\tau) =
\left\{
\begin{array}{lr}
\arcsin^{2}\sqrt{1/ \tau}\,,&\tau\geq 1\\
-\dfrac{1}{4}\left[\log\dfrac{1+\sqrt{1-\tau}}{1-\sqrt{1-\tau}}-i\pi\right]^{2}\,, \qquad &\tau< 1
\end{array}
\right.
 \end{equation}

Using the above definitions in the decay width formula given in Eq.~(\ref{decay_rate}), we arrive at a much simplified expression for the decay width,
\begin{equation}
\Gamma(h\rightarrow \gamma\gamma)=\frac{G_{\mu}\alpha^{2}m_{h}^{3}}{128\sqrt{2}\pi^{3}}
\left\vert g_{hVV} A^{h}_{W}+\frac{4}{3}g_{htt}A^{h}_{t}
\pm \frac{1}{3}g_{hbb}A^{h}_{b}+ \kappa A^{h}_{\xi}
\right\vert^{2}\,,
\label{decay_rate_simplified}
\end{equation}
where the $'+'$ sign before $A^{h}_{b}$ is for when the $h\bar bb$ Yukawa coupling has the same sign as the $hVV$ coupling 
and the $'-'$ sign is for the wrong sign of the Yukawa coupling, and $\kappa$ is defined as
\begin{equation}
\kappa=\frac{1}{m_{\xi}^{2}}(m_{\xi}^{2} + \frac{1}{2}m_{h}^{2} - m_{A}^{2})\,.
\label{Kappa_def}
\end{equation}

The appearance of $m_{A}$ in Eq.~(\ref{Kappa_def}) is merely an artefact of U(1) symmetry of the scalar potential. For a more general potential the expression for $\kappa$ involves 
$\lambda_{5}$~\cite{Arhrib:2003ph}.
\begin{figure}[htbp]
	\begin{subfigure}{
			\includegraphics[height=0.3\columnwidth, width = 0.45\columnwidth]{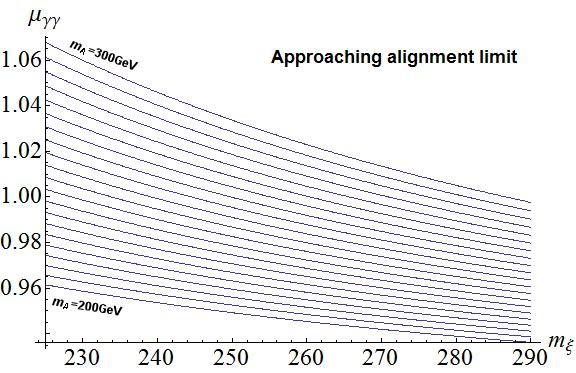} 
			\label{diphoton_align.fig}}		
	\end{subfigure}
	\begin{subfigure}
		{	\includegraphics[height=0.3\columnwidth, width = 0.45\columnwidth]{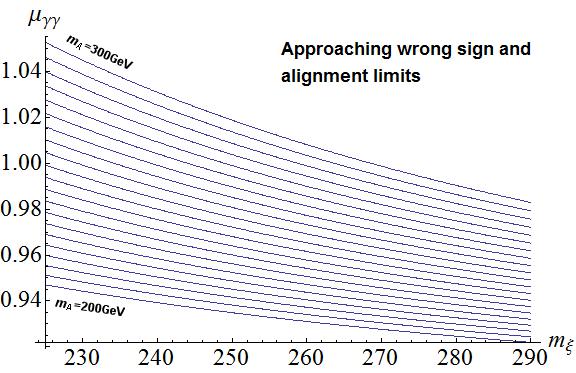} 
			\label{diphoton_WSL.fig}}     
	\end{subfigure}
	\caption{\small Diphoton decay width of the SM-like Higgs particle (normalized to SM) 
		as a function of the charged Higgs mass in GeV at $\tan \beta = 17$\,, for (a) same sign and (b) wrong sign, of
		down-type Yukawa couplings. }
	\label{diphoton.fig}
\end{figure}
%
In Fig.~\ref{diphoton.fig} we have plotted the $h\to\gamma\gamma$ decay width in 2HDMs in the alignment limit,
normalized with respect to the SM value, against the mass of the charged Higgs particle,
and for different values of the mass of the CP-odd scalar. 
Fig.~\ref{diphoton_align.fig} shows the decay width for the case where the 
$h\bar{q}q$ Yukawa coupling has the same sign as the 
$hVV$ coupling, whereas Fig.~\ref{diphoton_WSL.fig} is for the decay width corresponding to 
the case where the Yukawa coupling of $h$ to the down-type quarks is of the opposite sign
to the $hVV$ coupling. We note that the first case has been plotted, albeit for smaller
values of $\tan\beta\,$ and without the use of the Veltman conditions (thus for a much larger
range of $m_\xi$), in~\cite{Bhattacharyya:2013rya}.

As we have seen in the previous section, simultaneously choosing the alignment limit 
and the wrong sign limit also sets $\tan\beta$ at a high value. The critical value 
$\tan\beta=17$\,, and a small but non-zero value of 
$\alpha\,,$ namely $\alpha\simeq 0.035\,,$ was chosen for both the plots. 
The plots are not noticeably different for other high values of $\tan\beta\,$ or other 
similar values of $\alpha\,.$
The decay width does not depend on the type of 2HDM once the masses of the charged Higgs 
particle and the CP-odd Higgs particle are fixed. However, the range of allowed masses depends 
on the type of 2HDM being considered. We have chosen the ranges 
225 GeV$\leq m_{\xi} \leq$290 GeV and 200 GeV$\leq m_A \leq$ 300 GeV which cover the 
allowed ranges for all four types for $\tan\beta = 17\,.$ Although a picture is worth a thousand words,
it is perhaps worth pointing out that when $m_A$ is small, for example $m_A \simeq 200$ GeV, 
the diphoton decay width deviates from the SM value by 5-7\% for all values of $m_\xi\,.$ The deviation
is noticeable for many other values of $m_A$ also, as can be easily seen from the plots. On 
the other hand, for specific choices of $(m_A\,, m_\xi\,)$ the $h\to\gamma\gamma$ decay width
is the same as for the SM, so the non-observation of a deviation does not rule out 2HDMs.

\begin{figure}[htbp]
\includegraphics[height=0.3\columnwidth, width = 0.45\columnwidth]{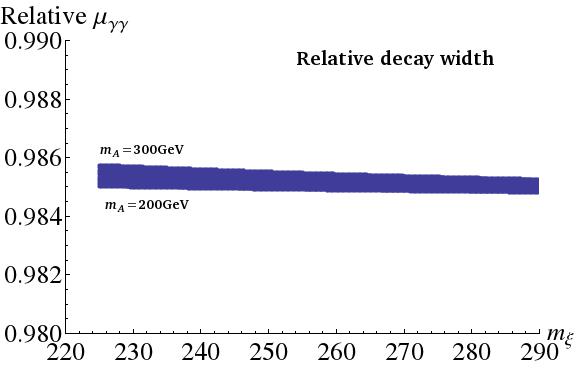} 
\caption{$h\gamma\gamma$ decay width for `wrong sign' $h\bar{D}D$ coupling relative to the
	case with `same sign' Yukawa couplings}
\label{relative.fig}
\end{figure}

The two plots are similar, but not identical. The decay width when the $h\bar{D}D$ 
Yukawa coupling is of the `wrong sign'  is smaller than the decay width for the
case when it is of the same sign (as $hVV$ couplings) by about 1.5\%, as can be
seen from the ratio of the decay widths, displayed in Fig.~\ref{relative.fig}.

\section{Results and Conclusion}
In this paper we have looked at how a certain criterion of naturalness, namely 
the cancellation of quadratic divergences, affect the allowed ranges of masses
of the additional scalars in 2HDMs in the alignment or SM-like limit with `wrong sign'
Yukawa couplings, and also in the reverse alignment limit. A similar calculation 
was done in~\cite{Biswas:2014uba} for the alignment limit without the 
`wrong sign' assumption.

We found that reverse alignment, \textit{i.e.} the scenario in which the heavier 
CP-even neutral scalar is the Standard Model Higgs particle, is clearly not a viable 
scenario for 2HDMs. Constraints 
arising from naturalness, stability, perturbative unitarity and experimental bounds
on the $\rho$-parameter completely rule out this scenario. The naturalness 
criterion is crucial for this conclusion -- reverse alignment is an allowed scenario if
quadratic divergences are taken care of by some mechanism of fine tuning, for example.

We have also considered a limit where the lighter CP-even neutral scalar corresponds to
the SM-like Higgs but where the Yukawa couplings of this particle to $D$-type quarks 
are of the wrong sign relative to their gauge couplings. In this scenario we obtain 
mass ranges for the rest of the physical Higgs bosons for various benchmark values of 
$\tan\beta$. In this paper we have shown only the plot for Type II 2HDM, but the results 
are similar for the other 2HDMs with a small variation of a few GeV.

The Higgs-diphoton decay width in a 2HDM receives additional contributions 
from loops containing the charged scalar $\xi^{\pm}$\,, so the decay 
width in a 2HDM is different from the SM value. Further, in the wrong sign limit,
loops containing down type quarks contribute with a different sign. 
We have plotted the $h\to 2\gamma$ decay width against 
the mass of the charged Higgs, and also for different values of
the mass of the CP-odd neutral scalar, and found that the decay width
can differ from its SM value by up tp 6\% for some values of the parameters.

While this paper was being completed, another paper which investigates 
what we call the reverse alignment limit appeared as an 
e-print~\cite{Bernon:2015wef}. However, that paper uses fewer constraints, 
so limits on the masses of $\xi^\pm$ are less restrictive.

More recently, the ATLAS and CMS collaborations at the LHC have reported an
excess corresponding to a diphoton resonance at 750 GeV~\cite{750GeV}. We note
that according to the naturalness criterion we have used in this paper, this
excess cannot be one of the scalar particles in any of the four types of
2HDMs, in agreement with the negative result found in~\cite{Angelescu:2015uiz} 
using several other lines of argument.  

\section*{Acknowledgement}
AB thanks Dipankar Das for useful discussions. The authors thank the anonymous referee for 
raising questions about the plot of the diphoton decay width in the first version, which 
helped us find and correct a mistake in the original plot. 

\end{document}